\documentclass{PoS}

\usepackage{amsmath,amstext,amssymb,bm,comment,color,graphicx,subfigure}

\newcommand{\C}{C}

\newcommand{\sm}{\theta}
\newcommand{\nn}{{\nonumber}}
\newcommand{\ddd}{\displaystyle}

\newcommand{\vektor}{A}

\newcommand{\gluon}{G}

\title{Chiral Superfluidity for the Heavy Ion Collisions}

\ShortTitle{Chiral Superfluidity for the Heavy Ion Collisions}

\author{\speaker{T. Kalaydzhyan}\\
        DESY Hamburg, Theory Group, Notkestrasse 85, D-22607 Hamburg, Germany\\
        ITEP, B. Cheremushkinskaya str. 25, Moscow, 117218 Russia\\
        E-mail: \email{tigran.kalaydzhyan@desy.de}}

\abstract{We argue that the strongly coupled quark-gluon plasma formed at LHC and RHIC can be considered as a chiral superfluid. The ``normal'' component of the fluid is the thermalized matter in common sense, while the ``superfluid'' part consists of long wavelength (chiral) fermionic states moving independently. We use the bosonization procedure with a finite cut-off and obtain a dynamical axion-like field out of the chiral fermionic modes. Then we use relativistic hydrodynamics for macroscopic description of the effective theory obtained after the bosonization. Finally, solving the hydrodynamic equations in gradient expansion, we find that in the presence of external electromagnetic fields the motion of the ``superfluid'' component gives rise to the chiral magnetic, chiral electric and dipole wave effects. Latter two effects are specific for a two-component fluid, which provides us with crucial experimental tests of the model. By considering probe quarks one can show that the fermionic spectrum at the intermediate temperatures ($T_c < T < 2 T_c$) has a gap between near-zero modes and the bulk of the spectrum - one more hint supporting the two-component model.}

\FullConference{Xth Quark Confinement and the Hadron Spectrum\\
                 8�-12 October 2012\\
                 TUM Campus Garching, Munich, Germany}

\begin{document}

\section{Introduction}

The non-trivial structure of the QCD vacuum attracted much attention in light of recent heavy-ion experiments performed at RHIC and LHC. These experiments provides us a possibility to study the strongly-coupled quark-gluon plasma (sQGP) in hadronic scale magnetic fields \cite{Tuchin:2013ie}. The non-trivial gluonic configurations may induce an imbalance between densities of left- and right-handed light quarks (chirality). As a strong magnetic field is applied to the system, the imbalance can give rise to a net electric current in sQGP along the magnetic field (chiral magnetic effect \cite{Kharzeev:2007jp, CME}). So far it was difficult to build a first-principles theory, describing this and similar effects, since the physics of sQGP is essentially nonperturbative. However, it seems that such a theory can be established (and it is sketched below), because QCD contains a long-wave axion-like degree of freedom, which can play a role of carrier for the chirality. Indeed, one can consider QCD coupled to QED, gauge $U_A(1)$ and bosonize quarks with Dirac eigenvalues smaller than some fixed $\Lambda$. As the result of such procedure one obtains the following effective Euclidean Lagrangian \cite{kalaydzhyan}
	        \begin{align}
		{\cal L}^{(4)}_E &= \frac{1}{4}G^{a\mu\nu}G_{\mu\nu}^{a} + \frac{1}{4}F^{\mu\nu}F_{\mu\nu} -j^{\mu} \vektor_\mu - g j^{a\,\mu} G^a_\mu \nn\\
 		& + \ddd\frac{\Lambda^2 N_c}{4\pi^2} \partial^\mu\sm\partial_\mu\sm\ +\frac{g^{2}}{16\pi^{2}}\sm\gluon^{a\mu\nu}\widetilde{\gluon}{}_{\mu\nu}^{a} + \frac{N_c}{8\pi^{2}}\sm F^{\mu\nu}\widetilde{F}{}_{\mu\nu}\\\label{lagrangian}
 		& +\ddd\frac{N_c}{24\pi^2}\sm \Box^2 \sm - \frac{N_c}{12\pi^2} \left( \partial^\mu\sm\partial_\mu\sm\right)^2\,,\nn
		\end{align}
where the axion-like field $\sm$ originates from the fermionic IR modes, while source currents can be formed by the UV modes.
The physical value for the cut-off $\Lambda$ is different in different regimes and can be found from the dependence of the chirality on the axial chemical potential \cite{kalaydzhyan}: $ \Lambda = \pi\sqrt{\frac{2}{3}}\sqrt{T^2 + \frac{\mu^2}{\pi^2}}$ at high temperatures; $\ddd\Lambda = 2\sqrt{|eB|}$ at strong magnetic fields; $\ddd\Lambda \simeq 3\, \mathrm{GeV}$ at small temperatures and weak (or absent) magnetic fields.

\section{Chiral superfluidity}
Chiral superfluidity is a hydrodynamic model of the sQGP in the range of temperatures $T_c < T \lesssim 2\,T_c$ derived from the effective Lagrangian (\ref{lagrangian}) (see \cite{kalaydzhyan} for details).
The essential idea of the model is to consider the sQGP as a two-component fluid with two independent motions: a curl-free motion of the near-zero quark modes (the ``superfluid'' component); and the thermalized medium of dressed quarks (the ``normal'' component). The first one is described by a pseudo-scalar
 field $\sm$, while the second one is represented by a four-velocity $u^\mu$.
Hydrodynamic equations 
 can be written in the following way \footnote{here we consider only
 color-singlet currents and neglect the anomaly induced by gluonic fields for simplicity. For the complete treatment see \cite{kalaydzhyan}}
\begin{align}
\partial_{\mu}&T^{\mu\nu} = F^{\nu\lambda} J_{\lambda}\,,\\
\partial_\mu& J^\mu = 0\,,\\
\partial_\mu& J_5^\mu = -\frac{\C}{4}  F^{\mu\nu} \widetilde F_{\mu\nu}\,,\\
u^\mu& \partial_\mu \sm + \mu_5 = 0\,,\label{Josephson}
\end{align}
where the constitutive relations are given by
\begin{align}
 T^{\mu\nu} &= \left(\epsilon + P\right)u^\mu u^\nu + P g^{\mu\nu} + f^2 \partial^\mu \theta\partial^\nu \theta+ \tau^{\mu\nu} \,,\label{cons1}\\
 J^\mu & \equiv j^\mu + \Delta j^\mu  = \rho u^\mu + \C \widetilde{F}^{\mu\kappa} \partial_\kappa \sm + \nu^\mu \,,\label{cons2}\\
 J_5^\mu &=  f^2 \partial^\mu \theta + \nu_5^\mu\,.\label{cons3}
\end{align}
Here $T^{\mu\nu}$ is the energy-momentum tensor of the liquid (total energy-momentum tensor minus the ones of free vector fields), $J^\mu$ is the total electric current, $J_5^\mu$ is the axial current.

The listed equations are similar to ones of a relativistic superfluid \cite{Son2000}, suggesting the name ``chiral superfluid'' for our model. Eq.~(\ref{Josephson}) is the Josephson equation. One should, of course, distinguish our situation from the conventional superfluidity as, first, there is no spontaneously broken symmetry and, second, the normal component is axially neutral.

Dissipative corrections $\tau^{\mu\nu}$, $\nu^\mu$, $\nu_5^\mu$ are due to finite viscosity, electric resistivity, etc., and do not contain terms proportional to the chiral anomaly coefficient $\C$ \cite{kalaydzhyan}.
Additional electric current $\Delta j_\lambda \equiv \C \widetilde{F}_{\lambda\kappa} \partial^\kappa \sm$ is induced by the ``superfluid'' component and can be represented by a sum of three terms
\begin{align}
\Delta j_\lambda = - \C \mu_5 B_\lambda + \C \epsilon_{\lambda\alpha\kappa\beta}u^\alpha \partial^\kappa \sm E^\beta - \C u_\lambda (\partial \sm \cdot B)\,,\label{phenom}
\end{align}
Where the terms correspond to the chiral magnetic effect \cite{Kharzeev:2007jp,CME}, to chiral electric effect \cite{CEE} and to the chiral dipole effect (generalization of the chiral magnetic wave \cite{CMW}), respectively.
Here the electric and magnetic fields in the ``normal'' rest frame are defined as
\begin{align}
\label{EB}
 E^\mu = F^{\mu\nu}u_\nu, \qquad B^\mu = \tilde F^{\mu\nu}u_\nu \equiv  \frac{1}{2}\epsilon^{\mu\nu\alpha\beta}u_\nu F_{\alpha\beta}.
\end{align}

By considering probe quarks one can show that the fermionic spectrum at the intermediate temperatures in interest ($T_c < T < 2 T_c$) consists of near-zero modes and the bulk of the spectrum, separated from the former by a gap. This gives us one more hint supporting the two-component model.
So far we are not able to identify the microscopic structure of the collective field $\sm$. One of the possibilities might be the long-distance propagation of light quarks along the low-dimensional extended structures, populating the QCD vacuum (see, e.g. \cite{Zakharov:2012vv, Chernodub:2012mu, kalaydzhyan} and refs. therein), if the structures exist beyond the probe quark limit.

\end{document}